\documentclass[preprint,authoryear]{elsarticle}
\usepackage{graphicx}  
\usepackage{txfonts}

\begin{document}

\title{Studies on the time response distribution of  \textit{Insight}-HXMT/LE}

\author[a,b]{Xiao-Fan Zhao}
\ead{zhaoxf@ihep.ac.cn}

\author[a,c]{Yu-Xuan Zhu}
\ead{zhuyx@ihep.ac.cn}

\author[a]{Da-Wei Han}
\author[a]{Wei-Wei Cui}
\author[a]{Wei Li}
\author[a]{Juan Wang}
\author[a]{Yu-Sa Wang}
\author[a]{Yi Zhang}
\author[a]{Yan-Ji Yang}
\author[a]{Bo Lu}
% ...

\address[a]{Key Laboratory of Particle Astrophysics, Institute of High Energy Physics, Chinese Academy of Sciences, Beijing 100049, China}
\address[b]{University of Chinese Academy of Sciences, Beijing 100049, China}
\address[c]{College of Physics, Jilin University, Changchun 130012, China}
\address[d]{Department of Astronomy, Beijing Normal University, Beijing 100088, China}
\address[e]{Department of Astronomy, Tsinghua University, Beijing 100084, China}
\address[f]{Department of Physics, Tsinghua University, Beijing 100084, China}
\address[g]{Department of Engineering Physics, Tsinghua University, Beijing 100084, China}
\address[h]{Key Laboratory of Space Astronomy and Technology, National Astronomical Observatories, Chinese Academy of Sciences, Beijing 100012, China}

\cortext[cor1]{Corresponding author}

\begin{abstract}
The Hard X-ray Modulation Telescope (HXMT) named \textit{Insight} is China's first X-ray astronomical satellite.
The Low Energy X-ray Telescope (LE) is one of its main payloads onboard.
The detectors of LE adopt swept charge device CCD236 with L-shaped transfer electrodes.
Charges in detection area are read out continuously along specific paths,
which leads to a time response distribution of photons readout time.
We designed a long exposure readout mode to measure the time response distribution.
In this mode, CCD236 firstly performs exposure without readout,
then all charges generated in preceding exposure phase are read out completely.
Through analysis of the photons readout time in this mode, we obtained the probability distribution of photons readout time.
\end{abstract}


\maketitle

\section{Introduction} 

The Hard X-ray Modulation Telescope \citep{Li+Wu+2008} is China's first X-ray astronomical satellite.
It was launched on June 15, 2017 and was named after \textit{Insight}.
The concept of \textit{Insight} was originally proposed in the 1990s,
which is based on the Direct Demodulation Method \citep{Li+Wu+1993, Li+Wu+1994}.
\textit{Insight} carries three main payloads:
High Energy X-ray Telescope (HE),
Medium Energy X-ray Telescope (ME)
and Low Energy X-ray Telescope (LE).
LE covers the energy band of 0.7--13\,keV
and uses detector arrays and collimators to collect X-ray photons \citep{Chen+etal+2018, Zhang+etal+2014}.
The swept charge device CCD236 is adopted as detector because of its high energy resolution and high time resolution.
The CCD236 is based on charge coupled device (CCD) technology,
and is electrically configured similar to a linear CCD \citep{Holland+Pool+2008}.
Charges generated in detection area are swept to the central readout amplifier continuously along specific paths,
so the position information of charges cannot be given \citep{Smith+etal+2015}.
Readout time of charges generated in different detection area are constant and related to the positions of charges,
which leads to a photon arrival time delay distribution --- time response distribution (TRD).

TRD has impact on the photon arrival time recorded by detector,
so the measurement of TRD is critical for timing analysis.
LE uses pulse profiles observed on the isolated pulsars to calibrate its timing system \citep{Zhang+etal+2018}.
Pulse profiles and the corresponding phase-resolved spectra help us explore pulsar radiation model.
TRD changes the shape of pulse profile and causes a phase delay \citep{Ge+2012}.
It is necessary to restore the pulse profile by TRD.
In addition, TRD is a significant part of getting time resolution of LE.

Long exposure readout mode (LERM) is a special readout mode of LE.
It can be used on the measurement of TRD.
Differing from normal readout mode (NRM), the readout of charges in detection area is not continuous.
CCD236 firstly performs exposure without readout for a specific period,
then all charges generated in the preceding exposure phase are read out completely.
This paper is focused on the measurement of TRD in LERM.
The following section introduces the readout method of CCD236 and the influences of collimators.
In Section \ref{Long exposure readout mode}, we describe LERM in detail.
Data analysis comes in Section \ref{Data analysis}.
In next section, we discuss the results.
At last we present a summary.

\section{Low Energy X-ray Telescope}
\label{Low Energy X-ray Telescope}

\subsection{Readout method of CCD236}

CCD236 is the second generation of swept charge device developed for X-ray spectroscopy.
It is designed for LE
and also used in India's Chandrayaan-2 Large Soft X-ray Spectrometer instrument
\citep{Radhakrishna+etal+2011, Smith+etal+2012}.
Comparing to the first generation swept charge device,
CCD236 benefits a lot from its significant improvements.
Firstly, CCD236 has a large detection area over 4\,cm$^2$.
Secondly, CCD236 reduces the drive phases from three to two,
which simplifies the clocking \citep{Holland+Pool+2008}.
Thirdly, CCD236 increases the element pitch from 0.02\,mm to 0.1\,mm,
which reduces charge transfer times of CCD236.
At last there is an improvement in radiation hardness \citep{Gow+etal+2009, Gow+etal+2015}.

\begin{figure}
\centering
\includegraphics[width=9cm, angle=0]{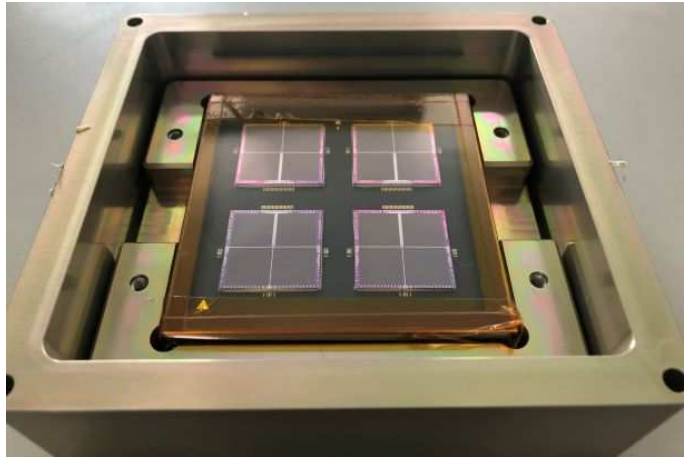}
\caption{CCD236 is divided into four quadrants
and four CCD236 detectors are packaged on an aluminum nitride ceramic substrate to make up a detector module.}
\label{Fig1}
\end{figure}

CCD236 is divided into four quadrants, and its output is located at central region.
In order to realize large-area detector arrays,
four CCD236 detectors are packaged on an aluminum nitride ceramic substrate to make up a detector module.
Figure \ref{Fig1} shows the detector module with four CCD236 detectors.

\begin{figure}
\centering
\includegraphics[width=6cm, angle=0]{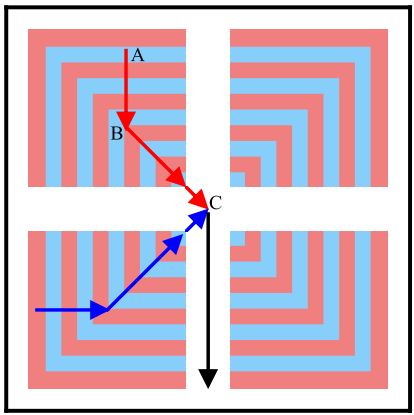}
\caption{Schematic of electron flow transfer paths.
Charge generated in position A is firstly transferred to position B along the direction perpendicular to the electrodes.
Then it is transferred to the output C along diagonal.
We call pixels in one L-shaped detection area as a ``packed-pixel''.
Charges in one packed-pixel need the same time to be read out.
The transfer paths of charges generated in other quadrants of CCD236 are similar, which is shown with blue lines.
}
\label{Fig2}
\end{figure}

To enable the suppression of surface dark current,
CCD236 is clocked continuously,
so the position information of charge signals cannot be given \citep{Smith+etal+2013}.
Because of its L-shaped transfer electrodes,
charges generated in detection area are firstly transferred to diagonal of CCD236 along the direction perpendicular to transfer electrodes.
Then charges are transferred to central output along the diagonal.
Charges generated in one L-shaped detection area need the same transfer time.
We call pixels in one L-shaped detection area as a ``packed-pixel''.
There are 100 packed-pixels per quadrant.
In addition, there are 20 pixels from central output to the L-shaped detection area.
For convenience, we also call these 20 pixels as packed-pixels,
so there are a total of 120 packed-pixels from central output to the edge of CCD236 in each quadrant.
Figure \ref{Fig2} gives schematic representation of electron flow transfer paths.

Among 120 packed-pixels, the area of about 20 packed-pixels near central output is pretty small.
The area of subsequent 98 packed-pixels increases gradually.
Due to the large noise,
charges generated in the outermost 2 packed-pixels near the edge of CCD236 are exported outside detector directly.
They are not involved in readout system.
In other words, the outermost 2 packed-pixels are designed as invalid pixels.

At the working frequency of 100\,kHz, all charges would be readout within 1.18\,ms.
Continuous readout mixes the charges from different packed-pixels,
so we are unable to tell the location and time of charge generation.
However, we could obtain the probability of different readout time statistically through TRD.

\subsection{The collimators and field of views}

\begin{figure}
\centering
\includegraphics[width=9cm, angle=0]{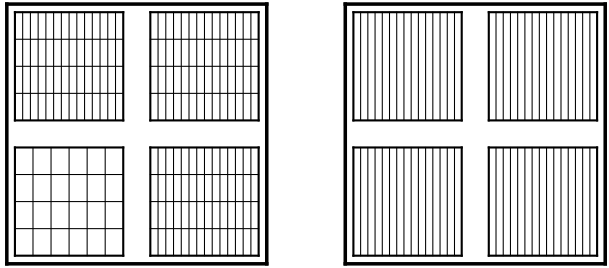}
\caption{The birds-eye schematic of long collimator with 3 narrow FOVs and a wide FOV (left), and the birds-eye schematic of short collimator with 4 ASM FOVs (right).}
\label{Fig3}
\end{figure}

LE combines a variety of field of views (FOVs) to accomplish scientific observation.
There is a collimator on top of each detector module to limit FOV.
The narrow FOV is $1.6^\circ\times6^\circ$, and the wide FOV is $4^\circ\times6^\circ$.
Besides these two FOVs, blocked FOV is added for the sake of better deducting background in both pointing observation and all-sky survey observation.
Blocked FOV does not collect X-ray photons from target source.
This paper is not involved in blocked FOV.
The three kinds of FOVs are limited by long collimator.
In addition, LE also designs a short collimator in order to monitor large-area sky in scanning observation,
which brings an all-sky monitoring (ASM) FOV. The ASM FOV is a trapezoidal FOV.
The birds-eye schematic of two kinds of collimators are shown in Figure \ref{Fig3}.

When observing an X-ray source, the collimator has an occlusion effect on CCD236,
especially for those pixels under collimators.
Considering the differences between TRD, which are caused by the structures of all kinds of collimators,
we need to treat different FOVs respectively.

\section{Long exposure readout mode}
\label{Long exposure readout mode}

In order to get the position information of charge signals, \citet{Smith+etal+2013} made several useful attempts.
One idea is to use an organic light emitting diode (OLED).
The OLED with a slightly larger area than CCD236 was placed in proximity with the CCD236 to provide a position-dependent signal.
The other is to use a X-ray beamline with 3\,mm spot size.
The CCD236 was scanned horizontally and vertically to provide a position-dependent signal.
By recording the position-dependent signal, \citet{Smith+etal+2013} obtained the response of CCD236 at different position.
Limited by the inability to get the pinpoint of single pixel, the results are relatively poor.
It is not necessary to obtain the time response of each single pixel.
LE designs long exposure readout mode to get TRD statistically.

There are two readout modes designed for LE: NRM and LERM.
NRM is the default readout mode.
In NRM, charges are transferred to output continuously without position information.
If the signal amplitude is higher than threshold,
an over threshold trigger would be recorded with its energy, arrival time and detector ID,
which is called a ``physical event''.
In addition, in order to monitor the signal baseline level of each detector,
the LE forcibly collects amplitude of each detector every 32\,ms in turn,
which is called a ``forced trigger event''.
LE box possesses 4 readout modules for 32 detectors in total, and every 8 detectors share the same readout module.
Forced trigger events have a higher priority than physical events.
When the forced trigger events are being recorded,
other events could not be read out for all detectors in the same readout module.
As a result, there are a portion of physical events that cannot be read out.

In LERM, charges are not transferred continuously.
CCD236 firstly performs exposure for a certain duration from several ms to several hundred ms,
during which the driving clock is paused,
and charges are stored in CCD236.
In the meantime, forced trigger events are not recorded either.
After that all charges in the chip will be continuously transferred and read out.
We design 150 clock cycles so that all charges generated in preceding exposure phase should be read out completely,
during which the readout of forced trigger events is resumed.

In continuous readout phase of LERM, charges stored in different packed-pixels are read out in sequence.
The one closest to output is read out firstly.
After 118 clock cycles, all charges in the chip are readout.
Nevertheless, new charges will generate during continuous readout phase,
because there still are incident X-ray photons meantime.
These photons are considered as out-of-time photons in LERM.

The main work frequency of LE is 100\,kHz.
At this work frequency, LE designs 7 kinds of exposure time: 2\,ms, 5\,ms, 10\,ms, 20\,ms, 50\,ms, 100\,ms, 200\,ms.
As the exposure time increases, the dark current of CCD236 increases.
This paper utilises the first 5 exposure time to measure time response distribution.

\section{Data analysis}
\label{Data analysis}

\subsection{Data set}

In ground experiments, we thoroughly test the long exposure readout mode of 5 kinds of exposure time in a vacuum chamber.
X-ray is generated mainly through fluorescence of vacuum chamber interior wall that is stimulated by the bremsstrahlung radiation from an X-ray tube,
hence the X-ray source is recognized as a diffused source.

During the periods of in-orbit operation, LE observed the source Swift J0243.6+6124 in 10\,ms long exposure readout mode.
Swift J0243.6+6124 is a Galactic ultraluminous X-ray pulsar discovered by \textit{Swift} \citep{Kennea+etal+2017},
and its rotation period is measured as 9.86\,s \citep{Kennea+etal+2017,Jenke+Wilson-Hodge+2017}.

\subsection{Split events reconstruction}

The split events are generated when an X-ray photon interacts with two or more adjacent pixels.
The criterion of split events for LE in NRM is that the arrival time interval of two or more events is not more than clock cycle
(0.01\,ms for 100\,kHz working frequency).
In LERM charges are stored in pixels before readout.
two or more neighboring single events may mistakenly be recognized as split events.
In this paper, we analyze the effect of split events reconstruction on the results.

\subsection{Epoch folding}

In order to obtain the X-ray photon counts of each clock cycle in readout phase, we perform an epoch folding.
The actual period of LERM has deviation from expectation.
For example, the whole period of 10\,ms long exposure is supposed to be 11.5\,ms (10\,ms exposure time plus 1.5\,ms readout time).
Nevertheless, actual period is about 11.4992 -- 11.4997\,ms and is time-dependent.
Charges in pixels are driven by the clock of LE box,
but their arrival time information of events is provided by the clock of satellite platform.
The two clocks cannot be completely consistent due to temperature drift.
After epoch folding, the TRD would have a large deviation from actual TRD.

In consideration of the periodicity of forced trigger events driven by the clock of LE box,
we recognize arrival time of forced trigger events as a time ruler to scale arrival time of physical events,
through which we solved the problem of clock inconsistency.

\subsection{Correction for forced trigger events}

\begin{figure}
\centering
\includegraphics[width=14cm, angle=0]{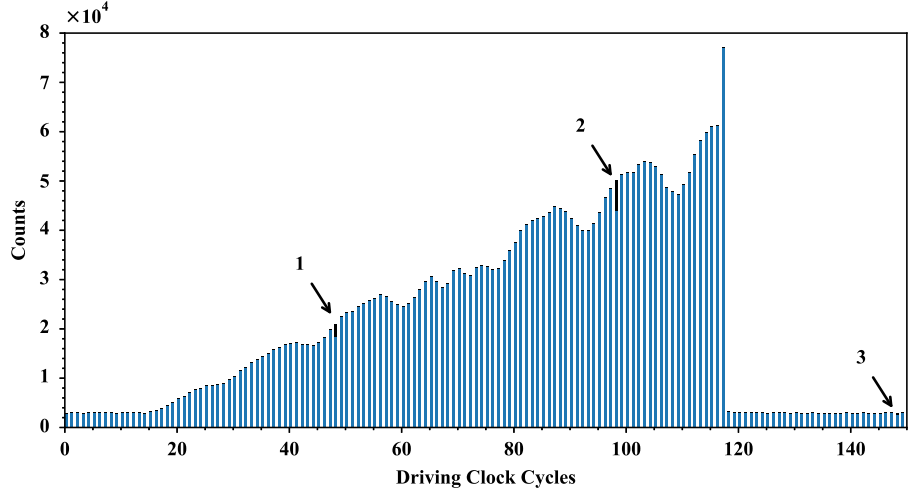}
\caption{X-ray photon counts of each driving clock cycle for narrow FOV in 10\,ms LERM. We take the result of narrow FOV in 10\,ms LERM as an example. The results of other FOV and other exposure time are similar. There are 3 positions where the counts are unusually lower than surrounding counts. The intervals between 3 positions are 50 clock cycles. These three black bars stand for event counts that are not read out because of forced trigger events.}
\label{Fig5}
\end{figure}

After epoch folding, we obtain the X-ray photon counts of each clock cycle in readout phase.
There are 3 unusually low counts due to the readout of forced trigger events.
In LERM, forced trigger events only appear in 3 fixed clock cycles and they have higher priority than physical events.
There is a certain proportion of time that physical events cannot be readout in these 3 specific clock cycles.
In the readout time of 150 clock cycles, an average of 0.5 forced trigger events will appear in each of these 3 positions.
When reading out forced trigger events, a total of 8 detectors cannot be read out,
accounting for a quarter of all detectors, as a result of which LE cannot read out physical events for one eighth of time.
Figure \ref{Fig5} shows the correction for forced trigger events.

\subsection{Correction for out-of-time events}

In LERM, the charges generated in exposure phase are what we need.
However, in the following readout phase there are still charges generated in pixels,
which are recognized as out-of-time events.
For charges in exposure phase, they are read out successively in first 118 clock cycles.
For those in readout phase, they are read out equiprobably in 150 clock cycles.
We use the average photon counts collected in the last 32 clock cycles as out-of-time photon counts.
After subtracting the out-of-time events, we finally obtain TRD.

From the results we could know that the counts of first 16 packed-pixels near output is so small that they can be ignored.
The counts of subsequent 102 packed-pixels increase gradually with a roughly linear growth.
This is in agreement with the detection area of each packed-pixel in CCD236.
In the trend of linear growth, there are several convex structures, which are caused by collimators.
Next section will describe the reason in detail.
The last packed-pixel is affected by outermost invalid packed-pixels, so its noise is relatively large.

\section{Results}
\label{Results}

\subsection{Comparison for split events reconstruction}

\begin{figure}
\centering
\includegraphics[width=9cm, angle=0]{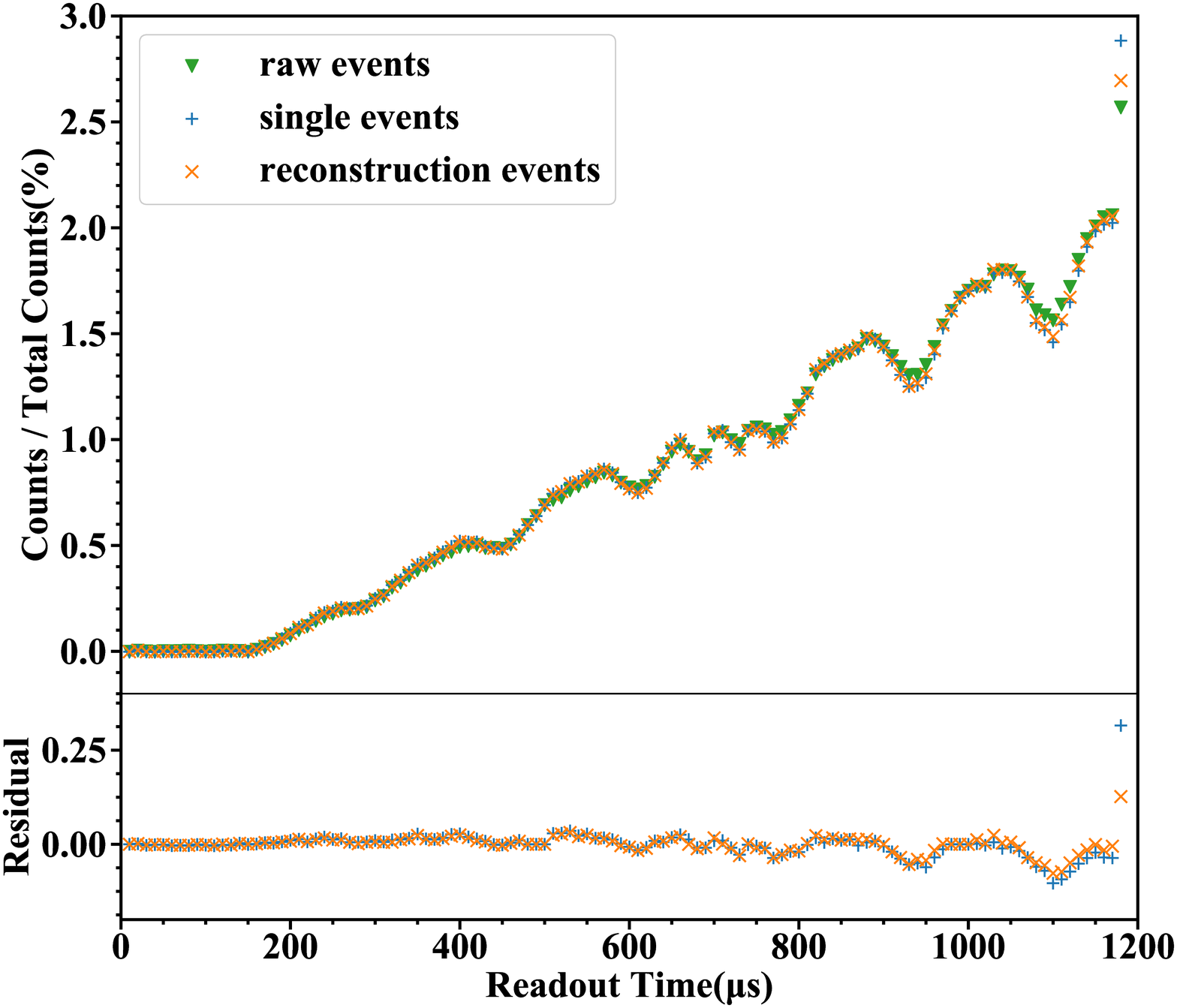}
\caption{TRD for narrow FOV in 10\,ms LERM. The statistical error is small, so we didn't draw it. Instead, we plot residual based on raw events. the X axis is transformed to readout time through driving clock cycle at 100\,kHz working frequency. The results of other FOV and other exposure time are similar.}
\label{Fig7}
\end{figure}

We compulsorily perform a split events reconstruction by the criterion of NRM.
Results are shown in Figure \ref{Fig7}.
Raw events stand for all events without reconstruction.
Single events stand for single events during split events reconstruction.
Reconstruction events are the sum of single events and split events.
The TRD for raw events, single events and reconstruction events are almost identical except the outermost packed-pixels shown in Figure \ref{Fig7}.
In the last packed-pixel, the ratio is the biggest for single events,
while the smallest for raw events.

When X-ray photons generate split events in the 118th and 119th packed-pixels,
charges in the 118th packed-pixel are recorded and those in 119th packed-pixel are discarded because of its unavailability.
After split events reconstruction, charges in the 118th packed-pixel are mistakenly recognized as single events,
which leads to the increment of ratio in single events.
In consideration that the difference is less than 0.5\%, this paper takes the results without split events reconstruction.

\subsection{Long exposure time}

\begin{figure}
\centering
\includegraphics[width=9cm, angle=0]{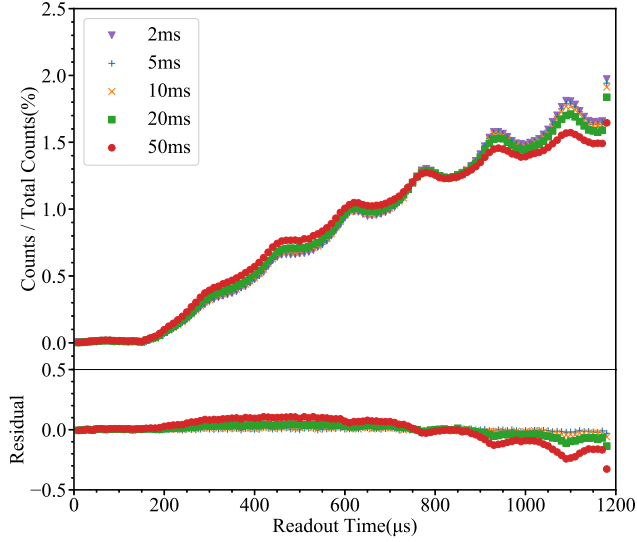}
\caption{TRD for ASM FOV in 5 kinds of exposure time. Residual is based on 2\,ms LERM. The results of other FOV are similar.}
\label{Fig8}
\end{figure}

During the periods of in-orbit operation, we only used 10\,mm exposure time to observe,
so we had to use the ground experiments to illustrate the effect of exposure time.
In ground experiments we tested 5 kinds of exposure time in LERM.
The results of ASM FOV are shown in Figure \ref{Fig8}.
As exposure time gradually increases,
the counts ratios of packed-pixels with smaller detection area also gradually increase, and those with larger detection area gradually decrease.

The pileup effect of LE is pretty small in NRM,
but in LERM the pileup effect is significantly higher than that in NRM.
Out-of-time events are read out continuously,
therefore the counting rate of out-of-time events stands for actual counting rate of X-ray photons.
We call the rest of events as long exposure events.
Table\ref{tab1} lists the counting rate of out-of-time events and long exposure events in ASM FOV for 5 kinds of exposure time.

\begin{table}
\begin{center}
\caption[]{Counting rate of out-of-time events and long exposure events for ASM FOV in 5 kinds of LERM.}
\label{tab1}

\begin{tabular}{cccc}
\hline\noalign{\smallskip}
No. &  exposure time      & Counting rate of out-of-time events & Counting rate of long exposure events  \\
\hline\noalign{\smallskip}
1  & 2\,ms     &   289.7$\pm{}$0.3\,counts s$^{-1}$     & 283.5$\pm{}$0.1\,counts s$^{-1}$  \\
2  & 5\,ms     &   290.6$\pm{}$0.4\,counts s$^{-1}$     & 276.9$\pm{}$0.1\,counts s$^{-1}$  \\
3  & 10\,ms   &   290.2$\pm{}$0.7\,counts s$^{-1}$     & 266.0$\pm{}$0.2\,counts s$^{-1}$  \\
4  & 20\,ms   &   289.3$\pm{}$0.7\,counts s$^{-1}$     & 246.4$\pm{}$0.1\,counts s$^{-1}$  \\
5  & 50\,ms   &   289.2$\pm{}$1.1\,counts s$^{-1}$     & 197.8$\pm{}$0.1\,counts s$^{-1}$  \\
\noalign{\smallskip}\hline
\end{tabular}
\end{center}
\end{table}

The X-ray intensities indicated by the counting rate of out-of-time events are the same in 5 kinds of long exposure experiments.
As the exposure time increases, the counting rate of long exposure events continues to decrease,
which indicates that the pileup effect is increasing.
The pileup effect becomes more significant as the pixel area increases,
as a result that the ratios of packed-pixels with larger detection area become smaller.

\subsection{The influence of FOV}

\begin{figure}
\centering
\includegraphics[width=9cm, angle=0]{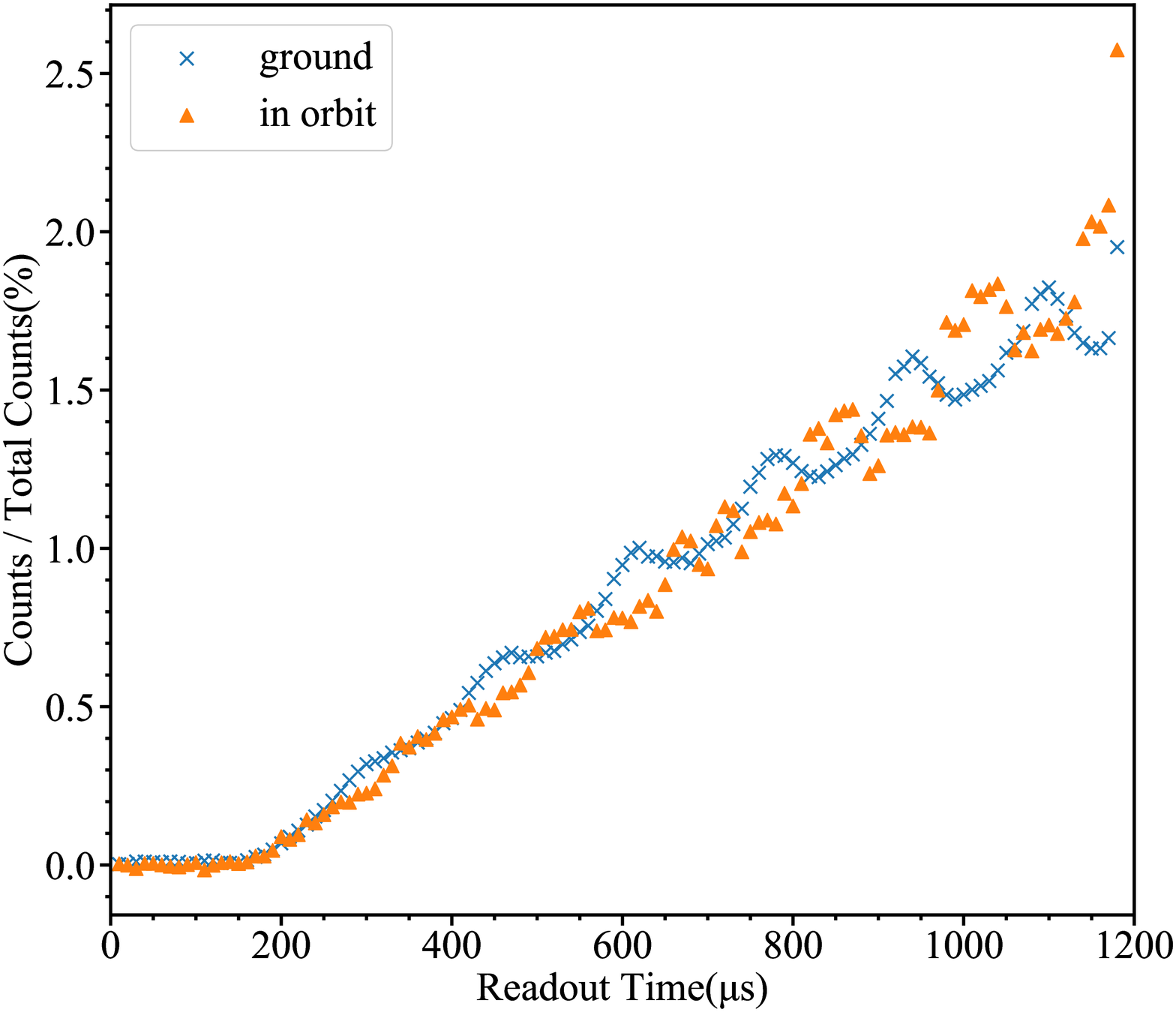}
\caption{TRD for 10\,ms ASM FOV in ground and in-orbit experiments.}
\label{Fig9}
\end{figure}

\begin{figure}
\centering
\includegraphics[width=9cm, angle=0]{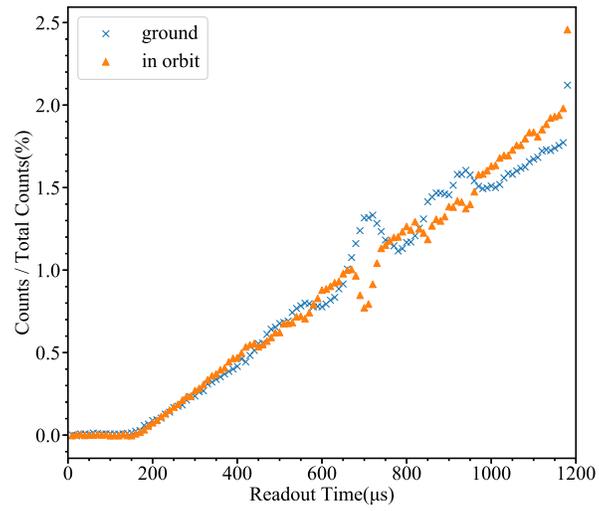}
\caption{TRD for 10\,ms wide FOV in ground and in-orbit experiments.}
\label{Fig10}
\end{figure}

\begin{figure}
\centering
\includegraphics[width=9cm, angle=0]{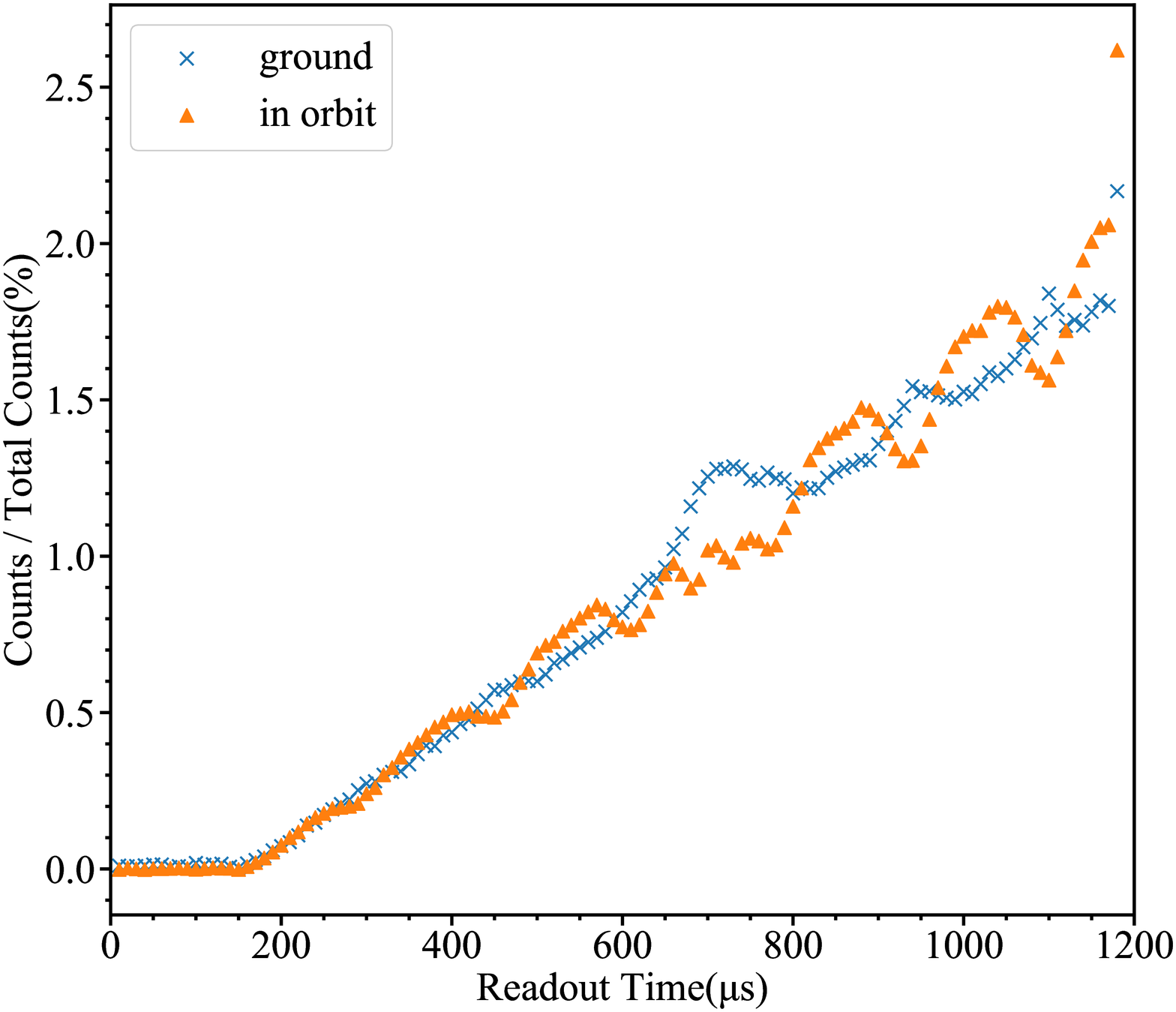}
\caption{TRD for 10\,ms narrow FOV in ground and in-orbit experiments.}
\label{Fig11}
\end{figure}

TRDs for ASM FOV, wide FOV and narrow FOV in ground and in-orbit experiments are respectively shown in Figure \ref{Fig9}, Figure \ref{Fig10}, Figure \ref{Fig11}.
In ground experiments there are several convex structures.
On the contrary, there are several concave structures in orbit.

ASM FOV, wide FOV and narrow FOV have 3 kinds of collimators shown in Figure \ref{Fig3}.
The collimators block X-ray photons and lead to special convex and concave structures in TRD.
Positions of collimators correspond with the convex and concave structures in TRD.

\begin{figure}
\centering
\includegraphics[width=9cm, angle=0]{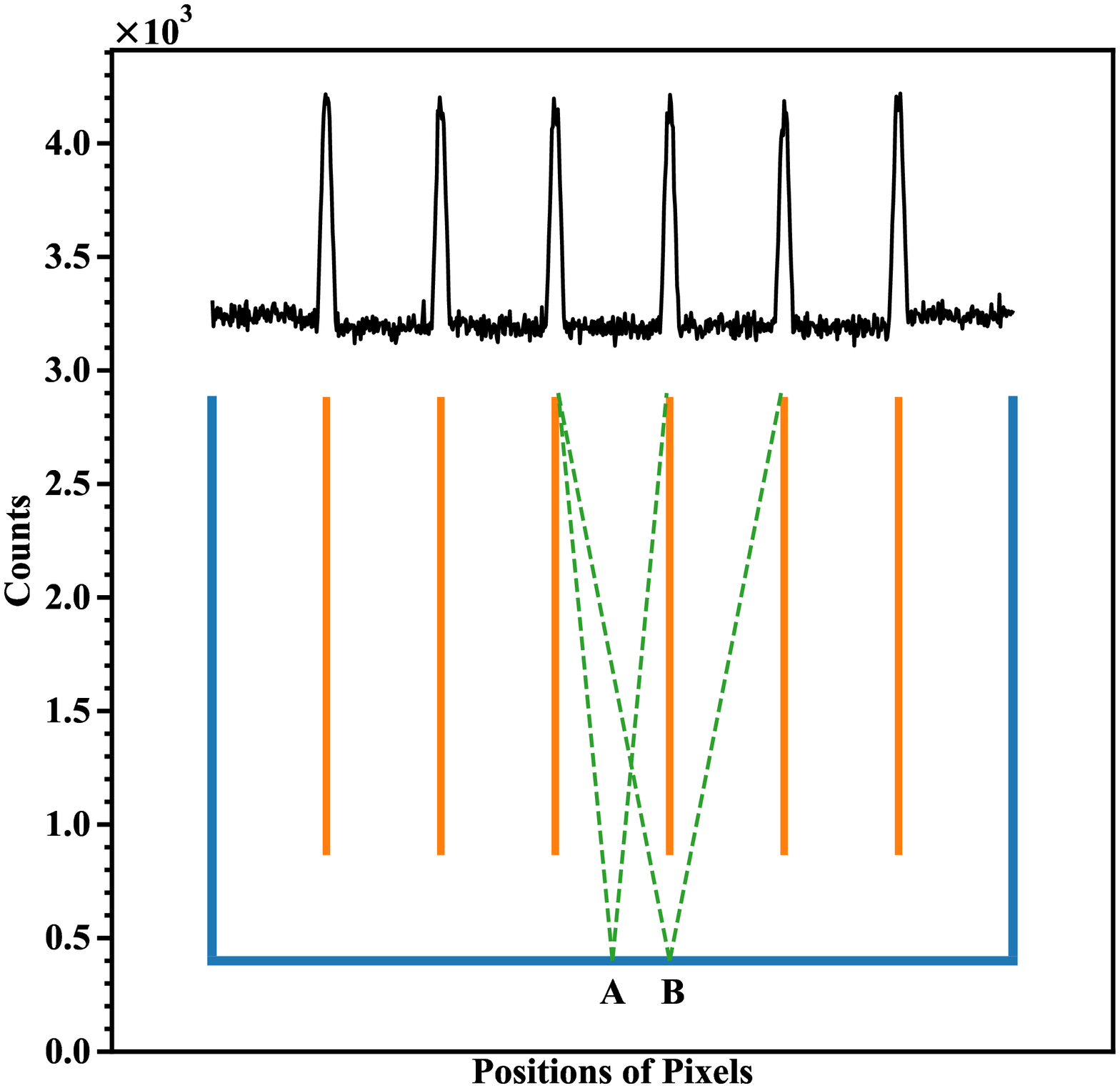}
\caption{The photon counts distribution under the illumination of a diffused source from simulation results.
Under the illumination of a diffused source, 
the photon counts from pixels under collimators is larger than those not under collimators.}
\label{Fig12}
\end{figure}

Ground observation and in-orbit observation present diametrically opposite results.
In ground experiments, the source is fluorescence of vacuum chamber interior which recognized as a diffuse source.
The angle of view from pixels under collimators is larger than that not under collimators.
We use a simplified one-dimensional model to simulate the effects of the collimators.
The result is shown in Figure \ref{Fig12}.
Under the illumination of a diffused source, 
the photon counts from pixels under collimators is larger than those not under collimators.
In contrast, the X-ray from Swift J0243.6+6124 in orbit is recognized as collimated X-ray.
The blocking of collimator should form a concave structure.

\section{Conclusions}

CCD236 is the second generation swept charge device with large photosensitive area.
Because of its L-shaped transfer electrodes and continuous readout, there is a TRD for photons arrival time.
We design long exposure readout mode to measure TRD in ground and in-orbit experiments.

The whole tendency of time response distribution is similar for ground and in-orbit experiments.
At the working frequency of 100\,kHz, all charges would be read out within 1.18\,ms.
The counts of events read out in first 16 driving clock cycles are so small that they can be neglected.
And counts of events in each driving clock cycle increase rough linearly for the following 102 driving clock cycles except several convex structures in ground experiments and concave structures in orbit.
The difference is caused by blocking of collimators when observing diffused and collimated X-ray sources.
With the increment of pileup effect, events count ratios for packed-pixels with smaller detection area increase and those with larger detection area decrease.

\section*{acknowledgements}
This work is supported by the National Natural Science Foundation of China under grants U1838201, U1838202.
This work made use of the data from the HXMT mission,
a project funded by China National Space Administration (CNSA) and the Chinese Academy of Sciences (CAS).

\end{document}